\documentclass[twocolumn,aps,prd,showpacs,amssymb,floatfix]
{revtex4}
\usepackage{mathptmx}      
\usepackage{bm}
\usepackage{amssymb,graphicx}

\begin{document}
\def\be{\begin{equation}}
\def\ee{\end{equation}}
\def\bea{\begin{eqnarray}}
\def\eea{\end{eqnarray}}
\def\ie{\hbox{\it i.e.}{ }}
\def\etc{\hbox{\it etc.}{ }}
\def\eg{\hbox{\it e.g.}{ }}
\def\cf{\hbox{\it cf.}{ }}
\def\viz{\hbox{\it viz.}{ }}

\title{Quark matter in neutron stars within the Nambu - Jona-Lasinio model
and confinement}

\author{M. Baldo, G. F. Burgio, P. Castorina, S. Plumari, and D. Zappal\`a}

\affiliation{Dipartimento di Fisica, Universit\`a di Catania
and INFN, Sezione di Catania, Via S. Sofia 64, I-95123, Catania, Italy} 

\date{\today}

\begin{abstract} 
The quark matter equation of state (EOS) derived from the standard 
Nambu - Jona-Lasinio (NJL) model is soft enough to render neutron stars (NS)
unstable 
at the onset of the deconfined phase, and no pure quark matter can be actually 
present in its interior. Since this is a peculiarity of the NJL model, 
we have studied a modified NJL model with a
momentum cut-off which depends on the density. This procedure, which
improves the agreement between QCD and NJL model at large density,
modifies the standard NJL equation of state, and then it is
potentially relevant for the stability analysis of neutron stars. 
We show that also within this approach, the NS instability
still persists, and that the vacuum pressure, as a signal of quark
confinement, has a fundamental role for the NS stability. In this
respect, our conclusions point to a  relationship between
confinement and NS stability.
\end{abstract}


\maketitle

\section{Introduction}
In the core of astrophysical compact objects, like neutron stars
(NS) or proto-neutron stars, nuclear matter is expected to reach a
density which is several times nuclear saturation density.
Calculations based on microscopic Equation of States (EoS), which
include only nucleonic degrees of freedom, show that the central
density of the most massive neutron stars can be from seven to ten
times the nuclear saturation density \cite{bbb}. In such
configurations the nucleons are closely packed, and to consider them
as separate entities becomes highly questionable. Unfortunately, it
is difficult to calculate accurately the transition point from
nucleonic to quark matter, and only rough estimates have been given
in the literature \cite{glu}. The microscopic theory of the
nucleonic Equation of States has reached a high degree of
sophistication, and different many-body methods have been developed.
They show a substantial agreement among each others \cite{tri}, and
therefore the main uncertainty of the transition point stays on the
quark matter EoS. Assuming a first order phase transition, as
suggested by lattice calculations, one can use different models for
quark matter to estimate the transition point, and calculate the
compact object configuration. This approach has been followed by
several authors, and a vast literature exists on this subject. The
applications of quark matter models to the study of the NS structure 
have used different versions of the MIT bag
model \cite{bag}, the color-dielectric model \cite{col} and 
different formulations of the NJL model
\cite{bub}. In general, it seems that the maximum mass of NS that
contain quark matter in their interior is bounded to be less than
1.6 - 1.7 solar masses. However, more recently it has been shown in
\cite{alf}  that if one corrects the MIT bag model by introducing
additional terms suggested by perturbative QCD, the maximum mass can
reach values close to 1.9 solar masses, similar to the ones obtained with
nucleonic degrees of freedom. Despite the similarity of the results
on the value of the maximum NS mass, the predictions on the NS
configurations can differ substantially from model to model. The
most striking difference is in the NS quark matter content, which
can be extremely large in the case of EoS related to the MIT bag
model or the color-dielectric model, but it is vanishingly small in
the case of the original version of the NJL model \cite{bub,sch}. In
the latter case it turns out that, as soon as quark matter appears
at increasing NS mass, the star becomes unstable towards collapse to
a black hole, with only the possibility of a small central region
with a mixed phase of nucleonic and quark matter. This result can be
quite relevant for the physics of NS, since the NJL model is the
only model which is based on phenomenological low energy data, i.e.
on hadron properties. The main drawback of the model is the absence
of confinement, since the gap equations which determine the quark
masses as a function of density cannot incorporate a confining
potential. One then assumes that the chiral phase transition marks
also the confinement transition, as indicated by all lattice
calculations. More recently a confining potential has been
introduced in the NJL model \cite{thom}, which is simply switched
off at the chiral phase transition. In this case indeed no sharp
instability of the NS is observed at the onset of quark matter in
the central core. This is suggestive of a connection between the
presence of confinement and the possibility of NS with a quark
matter central core. However it is not clear if the instability of
the NS is related or not to confinement for (at least) two reasons.
Indeed, the introduction of the confining potential requires several
parameters and, moreover, the NJL model at large density suffers of
another drawback pointed out in \cite{casa}. In fact, it has been
shown that the standard NJL model is not able to reproduce the
correct QCD behavior of the gap for large density, and therefore a
different cut-off procedure at large momenta has been proposed. More
precisely, a density dependent cut-off has been introduced, and this
strongly modifies the standard NJL model thermodynamics. Therefore,
the stability analysis of NS by the new EoS, which follows from this
modified treatment of NJL model at large density, is a preliminary
step to understand whether confinement is  an essential  ingredient
in  the stabilization mechanism.
\par
It is the purpose of this paper to clarify this point and to
identify the origin of the NS instability at the quark onset within
the original NJL model,  which sharply distinguishes this model from
all the others. We show that with the density dependent cut-off
procedure the NS instability still persists and that the vacuum
pressure, as signal of quark confinement, has a fundamental role for
the NS stability, as yet observed in the MIT bag model. In this
respect, our conclusions point to an indirect relationship between
confinement and NS stability.

\section{NJL at large density}

The NJL model provides a good phenomenological description of low energy QCD
based on the Lagrangian (for the two flavor case)
\be
{\cal L}_{NJL}=i\bar{\psi} \gamma^{\mu} \partial_{\mu} \psi +
g  \bigg[ \Big( \bar{\psi}  \psi \Big)^2
+  \sum_{\alpha=1}^{N_{f}^{2}-1}\Big( \bar{\psi} \tau^{\alpha}
i \gamma_5 \psi \Big)^2  \bigg]
\ee

As pointed out in \cite{casa}, the model is not able to reproduce the correct behavior of the
solution of QCD gap equation  at large density. Indeed, at zero temperature,
 the relevant dynamical degrees of freedom have momenta in the shell
$|p| \simeq \mu + \delta$, where $\mu$ is the
chemical potential and  $\delta$ is the cut-off evaluated from the Fermi momentum.
Therefore the standard  cut-off $\Lambda$ on the momentum,  $(|p| < \Lambda)$,
forces $\delta \simeq 0$ when $\mu \simeq \Lambda$, and this gives
an unphysical reduction of the number of dynamical states of the system.

To solve this problem a new procedure has been proposed, which
simultaneously  preserves the low energy properties of the theory,
and allows one to write a cut-off independent gap equation at zero
temperature and density.
The idea is the introduction of  a cut-off dependence in the four-fermion coupling constant, $g$,  by: \\
1)  keeping fixed  $f_\pi$ to its experimental value and  deriving the constituent mass $M$
as a function of $\Lambda$,  $M(\Lambda)$, from the expression for $f_{\pi}$:
\be
f_{\pi}=3M^2 \int \frac{d^3p}{(2\pi)^3} \frac{1}{E_p^3} \theta(\Lambda-|\vec{p}|)
\ee
2) determining the coupling $g(\Lambda)$ which solves the  gap equation, at $\mu=0$ and zero temperature,
\be
\label{gapeqstand}
M(\Lambda)=m+4 N_f N_c \: g(\Lambda)\int_{\Lambda} \frac{d^3p}{(2\pi)^3}
\frac{M(\Lambda)}{\sqrt{p^2+[M(\Lambda)]^2}}
\ee
for any value of $\Lambda$.
As shown in Fig. 1 of  \cite{casa}, $\Lambda^2 g(\Lambda)$  smoothly decreases with  $\Lambda$.

Then, according to the procedure developed in \cite{casa}, at finite density, a $\mu$ dependent cut-off
$\Lambda(\mu)$ is introduced, which, in turn, implies a $\mu$ dependent coupling constant.
In particular, a linear dependence in  $\Lambda(\mu)$ is taken, which  provides a better agreement
with high density QCD.

On the other hand this procedure  strongly changes the
thermodynamics and the EoS of the system at finite density with
respect to the standard NJL approach. In fact, the functional form
of the thermodynamical potential $\Omega$ is not modified by a $\mu$
dependent cut-off, and therefore the pressure is \be
p=-V^{-1}\Omega(\Lambda(\mu),g(\mu),\mu,T=0) \ee where $V$ indicates
the three-dimensional volume of the system and $\Omega$, in the
following, will be computed in the usual  mean field approximation.
The particular choice of $\Lambda(\mu)$ considered in  \cite{casa}:
\be \label{eq:cutoff} \Lambda(\mu)= \left \{ \begin{array}{ll}
\Lambda_0 & \quad \mu< {\Lambda_0}/{(c+1)}\approx 435~ {\rm MeV} \\
(c+1)\mu & \quad \mu> {\Lambda_0}/{(c+1)}\approx 435~ {\rm MeV}
\end{array} \right.
\ee
where $\Lambda_0\sim \rm 600~MeV$ is the  NJL cut-off in the standard treatment,
and $c=\delta/\mu = 0.35$ (see \cite{casa}),
introduces a discontinuity in the density $\rho=-\frac{1}{V}
(\partial\Omega/ \partial \mu)$,
and in the corresponding EoS. This definition of the density $\rho$
takes care of the dependence of $\Lambda$ on $\mu$, and ensures 
thermodynamical consistency. 

\begin{figure}[t]
\centering
\includegraphics[width=6.5truecm,angle=270]{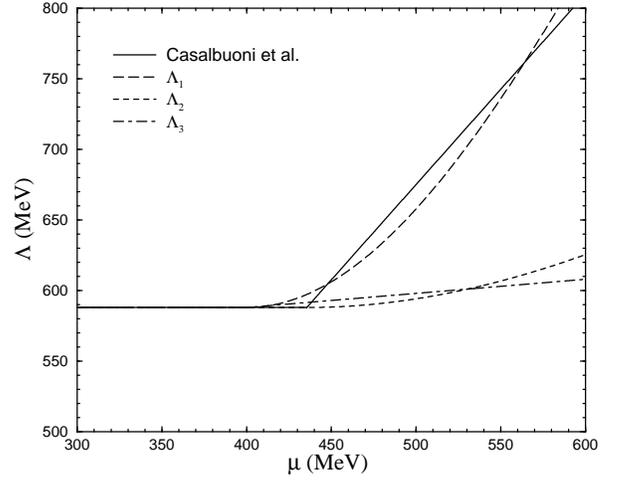}
\caption{The cut-off is displayed as a function of the chemical potential.
Several functional dependencies have been adopted, as discussed in the text.}
\label{f1}
\end{figure}

\begin{figure}[t]
\centering
\includegraphics[width=8.5truecm,angle=270]{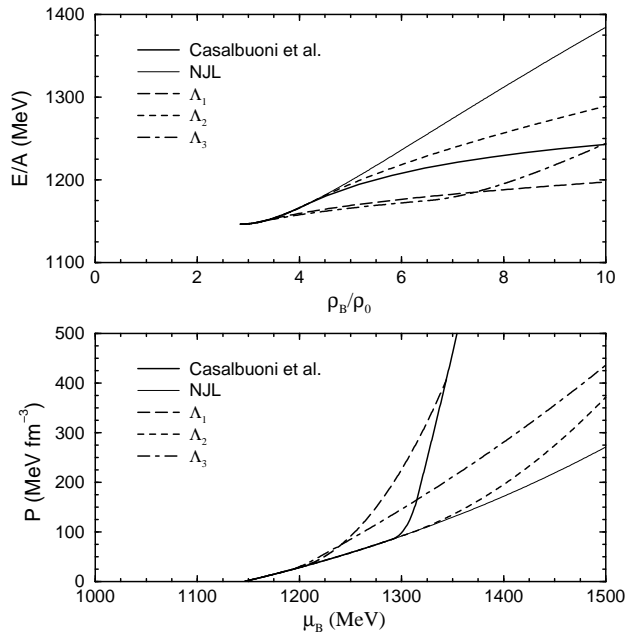}
\caption{The energy per baryon is shown as a function of the baryon
density (upper panel), whereas the pressure as a function of the
baryon chemical potential is displayed in the lower panel, for
different choices of the cut-off. See text for details.} \label{f2}
\end{figure}

Since there is no compelling restriction to a specific functional
form of $\Lambda(\mu)$, we replace Eq.(\ref{eq:cutoff}) with some
smooth interpolations that have no discontinuous densities and grow
with different slopes for large values of the chemical potential. In
Fig.1 we plot the cut-off of Eq. (\ref{eq:cutoff}) together with
three different ansatz:

\be \label{eq:cutoff1} \Lambda_1(\mu)= \left \{ \begin{array}{ll}
\Lambda_0 & \quad \mu \lesssim 400 ~ {\rm MeV} \\
\sqrt{9(\mu-400)^2+\Lambda_0^2} & \quad \mu \gtrsim 400~{\rm MeV}
\end{array} \right.
\ee

\be \label{eq:cutoff2} \Lambda_2(\mu)= \left \{ \begin{array}{ll}
\Lambda_0 &  \mu \lesssim 435~ {\rm MeV} \\
(c+1)\mu [{\rm log}(\frac{(c+1)\mu}{\Lambda_0})-1] +2\Lambda_0 & \mu
\gtrsim 435~{\rm MeV} 
\end{array} \right.
\ee

\be \label{eq:cutoff3} \Lambda_3(\mu)= \left \{ \begin{array}{ll}
\Lambda_0 & \quad \mu \lesssim 397~ {\rm MeV} \\
0.1 \mu + 548 & \quad \mu \gtrsim 403~{\rm MeV}
\end{array} \right.
\ee

\begin{figure}[t]
\centering
\includegraphics[width=6.5truecm,angle=270]{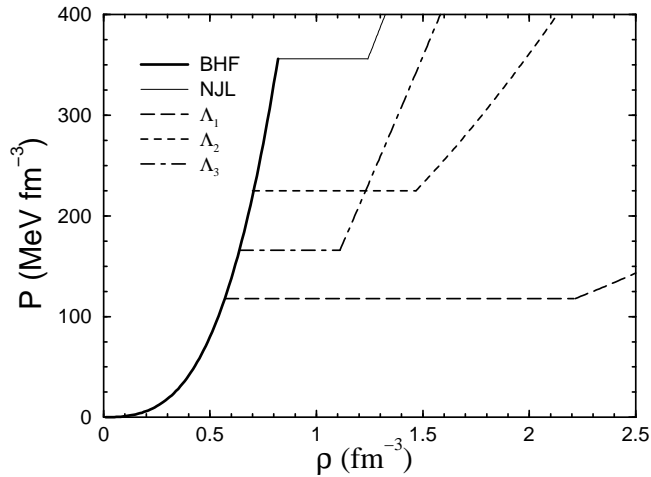}
\caption{The complete EOS is displayed, i.e. the pressure as function of the 
baryon density. See text for details.}
\label{f3}
\end{figure}

The numerical coefficients in Eqs. (\ref{eq:cutoff1}), (\ref{eq:cutoff2}), (\ref{eq:cutoff3}),
are adjusted in order to keep the curves in Fig.1 either close to the one of  Eq. (\ref{eq:cutoff}),
or close to the constant standard cut-off $\Lambda_0$. In particular, for  $\Lambda_3$,
in the interval  $397~{\rm MeV} < \mu <403~{\rm MeV}$,  a  continuous  quadratic interpolation between
the two straight lines reported in Eq. (\ref{eq:cutoff3}), is taken. We stress
that the adopted smoothing procedure is not intended to propose a method 
alternative to the one presented in \cite{casa}, but it has the only purpose 
of making this approach suitable for our application to NS, and avoiding 
spurious discontinuities in the density.

\begin{figure*}[t]
\centering
\includegraphics[width=8.truecm,angle=270]{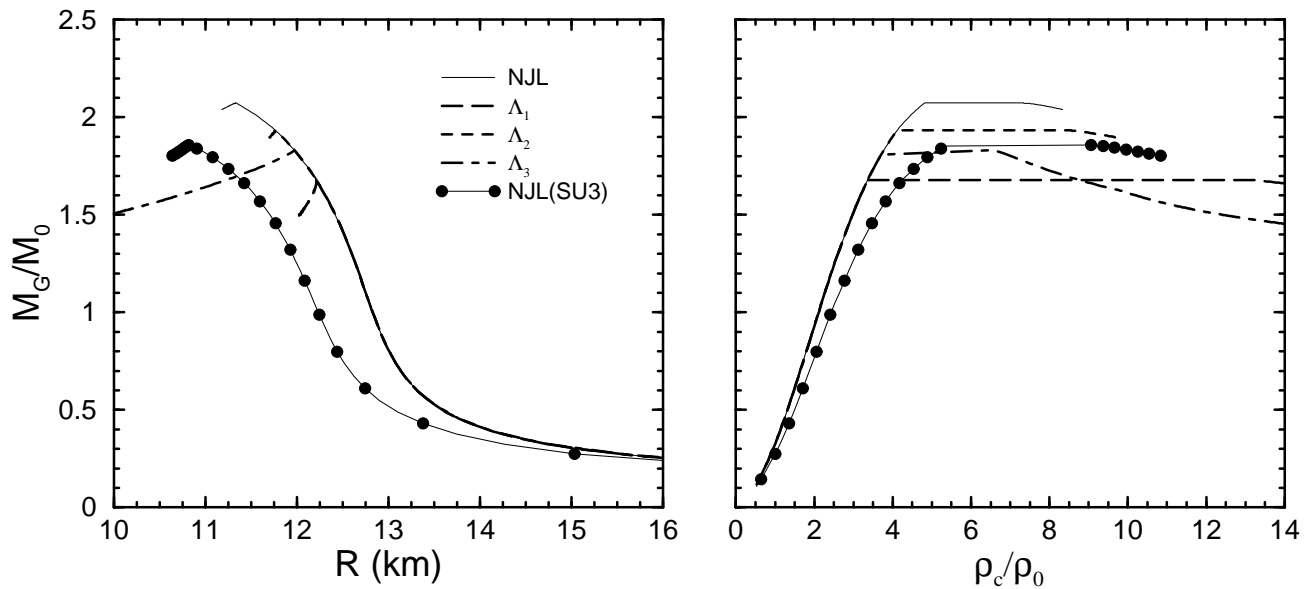}
\caption{The gravitational mass (in units of the solar mass
$M_\odot=2\times 10^{33}g$) is plotted as function of the radius (left panel)
and the central density (right panel), for different choices of the cut-off
behavior.}
\label{f4}
\end{figure*}

Considering different slopes in the $\mu$ dependence of the cut-off
is an important issue, because this implies different growth
behaviors of the pressure, and therefore of the density, as a
function of the baryon chemical potential $\mu_B$,
and moreover different dependence of the energy per baryon on the baryon
density $\rho_B$, which is crucial to determine the critical
value of $\rho_B$ in the transition from nuclear to quark matter.

In Fig.2  we display in the upper panel the energy per baryon $E/A$
versus $\rho_B / \rho_0$ for the pure quark phase 
($\rho_0$ indicates the nuclear saturation
density, $\rho_0=0.17~ \rm fm^{-3}$), and the pressure as a function of 
$\mu_B$ in the lower panel for the various $\Lambda_i (\mu)$. 

For completeness, we also
show the results obtained using a constant cut-off, as in the
standard NJL model (thin solid line), for which the parametrization
of ref.\cite{sch} has been used. As we shall see in the next
section, the stability analysis of the NS is influenced by the
various behavior of $E/A$ plotted in Fig. 2.

\section{NJL and Neutron Star stability}

The transition to quark matter in the core of NS has been studied by
many authors by matching the baryonic  EoS, as determined by
microscopic many-body calculations involving nucleons only, and the
quark EoS as determined by simple models which simulate QCD. The
crossing of the two EoS in the pressure vs. chemical potential plane
marks the transition point between the two phases. At the transition
the two phases are in mechanical and chemical equilibrium.  Below
the transition point, i.e. at lower density,  the nucleonic phase is
favored, while above the point the quark matter is favored. This
corresponds to the Maxwell construction for a first order phase
transition, as suggested by the indications coming from lattice
calculations \cite{fodor}. Solving the Tolmann-Oppenheimer-Volkoff 
(TOV) equations \cite{shap}, one can then calculate the NS
configuration and the corresponding gravitational mass as a function
of the central density or of the corresponding NS radius. In a
previous work \cite{bub}, where the NJL model was used, it was found
that the quark onset at the center of the NS as the mass increases
marks an instability of the star, i.e. the NS collapses to a black
hole at the transition point since the quark EoS is unable to
sustain the increasing central pressure due to gravity. For the
nucleonic sector a microscopic EoS derived with the
Brueckner-Bethe-Goldstone (BBG) method was used \cite{bbs}. 
The uncertainty in the
baryonic EoS is relatively small and not relevant to the present
analysis. In fact, the results are in line with similar calculations
within the NJL model, where other nucleonic EoS are used \cite{con}.
For simplicity here we restrict the calculations to the two flavor
case.
\begin{figure}[t]
\centering
\includegraphics[width=7.truecm,angle=270]{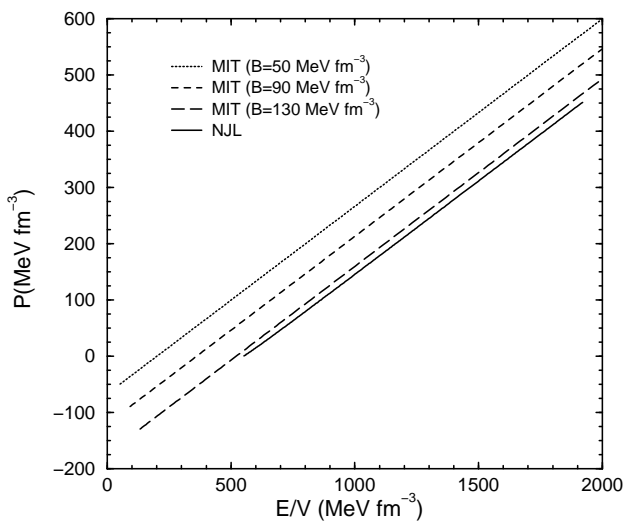}
\caption{The pressure is plotted as function of the energy density for the
two-flavour case.}
\label{f5}
\end{figure}

In the calculations presented in this paper, for the hadronic phase we 
have adopted a nucleonic equation of state obtained within the BBG approach
\cite{zuo}, using the Argonne $\rm v_{18}$ two-body potential \cite{wir},
supplemented by the Urbana phenomenological three-body force \cite{schi}.
For the quark phase, we use the standard parametrization of the NJL model 
\cite{sch}, and our proposed prescriptions of the cut-off described in Sect.II.
In Fig.3 we display the complete EOS, i.e. the pressure 
as function of the baryon density for the cases discussed above. The thick 
solid line displays the equation of the state for the pure hadron phase, 
treated in the BBG approach. The thin monotonically increasing curves
represent the quark matter branch of the EOS, whereas the plateaus are a 
consequence of the first order Maxwell construction.  
The transition density range shown by the plateau depends on the adopted 
cut-off procedure.

In Fig. 4 we report the neutron star masses as a function of the
radius (left panel), and of the central density (right panel) for
the different choices of the density dependent cut-off discussed
above. For comparison also the results for the standard NJL model
(three flavor) of ref. \cite{bub} are reported (full circles). In
all cases one can see that at the maximum mass the plot is
characterized by a cusp, which corresponds to the instability
mentioned above. The plateau which appears in each of the curves
reported in the right panel is a consequence of the Maxwell
construction. The width of the plateau is the jump of the central
density at the onset of quark matter.

It is likely that if
a mixed phase, according to the Glendenning construction \cite{glen},
had been used, then the plot would be smooth, but in any case no
pure quark matter is expected to be allowed in a stable NS. 

Therefore the improvement of the NJL model by the density  dependent
cut-off does not solve the NS stability problem, and the origin of
the instability should be related to the other missing dynamical
ingredient, i.e. the quark confinement.

To have an indication on the role of confinement, let us  consider
the  pressure as a function of energy density for the NJL model, as reported in
Fig.5. One can see that in the relevant range, since the chiral
symmetry is restored, the model behaves like the MIT bag model (i.e.
a free Fermi gas) with a bag constant $B_{NJL}$ close to about 140
MeV fm$^{-3}$. It is well known that at such value of the bag
constant the maximum mass within the MIT bag model is much smaller
than the value at the maximum of the curves in Fig.3. In general,
the maximum mass of NS increases at decreasing value of $B$
approximately as $B^{-1/2}$ \cite{wit}.

It looks likely that the same mechanism is acting in the three flavor case.
In fact we have seen that the new cut-off prescription 
$\Lambda=\Lambda(\mu)$ tends to soften the quark EOS with respect to the 
standard NJL. Since the latter displays already the NS instability discussed
above even with three quark flavors \cite{bub}, we can expect an enhancement
of the instability in the present case.
Within the
NJL model the value of the effective bag constant $B_{NJL}$ is dictated
by the low energy phenomenology and it cannot be tuned. Indeed,
the physical content of the model demands that the pressure in vacuum is zero,
since there is no confinement,  and then the constant pressure $B_{NJL}$
above the chiral transition
is necessarily present and determined uniquely by the parameters of the model
\cite{bubtes}. If one adds by hand a confining potential which is switched
off at the chiral phase transition \cite{thom}, then of course the
instability can be removed since the effective bag constant would be
correspondingly reduced.

In any case, this shows that the instability
is closely linked to the lack of confinement in the original NJL model.


\begin{thebibliography}{99}

\bibitem{bbb} M. Baldo, I. Bombaci, and G. F. Burgio, Astronomy and
Astrophysics {\bf 328}, 274 (1997).
\bibitem{glu} M. Baldo, P. Castorina and D. Zappala', Nucl. Phys. A
{\bf 743}, 44 (2004).
\bibitem{tri} M. Baldo, "Nuclear matter Equation of State at high
density", Proceedings of the 5th Conference on Hadron Physics at
ICTP, 2006, Nucl. Phys. A, C. Ciofi degli Atti and D. Treleani Eds,
to be published.
\bibitem{bag} G. F. Burgio, M. Baldo, P. K. Sahu, A. B. Santra, and
H.-J. Schulze, Phys. Lett. B {\bf 526}, 19 (2002); 
G. F. Burgio, M. Baldo, P. K. Sahu, and
H.-J. Schulze, Phys. Rev. C {\bf 66}, 025802 (2002), and references therein.
\bibitem{col} C. Maieron, M. Baldo, G. F. Burgio, and H.-J. Schulze, 
Phys. Rev. D {\bf 70}, 043010 (2004), and references therein.
\bibitem{bub} M. Baldo, M. Buballa, G. F. Burgio, F. Neumann, M. Oertel, and
H.-J. Schulze, Phys. Lett. B {\bf 562}, 153 (2003), and references therein.
\bibitem{alf} M. Alford, M. Brady, M. Paris and S. Reddy, ApJ {\bf 629},
969 (2005).
\bibitem{sch} K. Schertler, S. Leupold and J. Schaffner-Bielich, Phys.
Rev. C {\bf 60}, 025801 (1999).
\bibitem{thom} S. Lawley, W. Bentz and A.W. Thomas, J. Phys. G {\bf 32}, 667
(2006).
\bibitem{casa}
R. Casalbuoni, R. Gatto, G. Nardulli and M. Ruggieri, Phys. Rev. D
{\bf 68}, 034024 (2003).
\bibitem{fodor} Z. Fodor and S. D. Katz, JHEP {\bf 0404}, 050 (2004).
\bibitem{shap} S.L. Shapiro and S.A. Teukolsky,
 {\it Black Holes, White Dwarfs and Neutron Stars}
(Jhon Wiley and Sons, New York, 1983).
\bibitem{bbs} M. Baldo, G. F. Burgio, and H.-J. Schulze,
Phys. Rev. C {\bf 61}, 055801 (2000).
\bibitem{con} D. P. Menezes and C. Providencia, Phys. Rev. C {\bf 68},
035804 (2003). 
\bibitem{zuo} X. R. Zhou, G. F. Burgio, U. Lombardo,
H.-J. Schulze and W. Zuo, Phys. Rev. C {\bf 69}, 018801 (2004).
\bibitem{wir} R. B. Wiringa, V. G. J. Stoks, and R. Schiavilla,
Phys. Rev. C {\bf 51}, 38 (1995).
\bibitem{schi} J. Carlson, V. R. Pandharipande, and R. B. Wiringa, 
Nucl. Phys. {\bf A401}, 59 (1983); R. Schiavilla, V. R. Pandharipande,
and R. B. Wiringa, Nucl. Phys. {\bf A449}, 219 (1986).
\bibitem{glen} N. K. Glendenning, Phys. Rev. D {\bf 46}, 1274 (1992).
\bibitem{wit} E. Witten, Phys. Rev. D {\bf 30}, 272 (1984).
\bibitem{bubtes} M. Buballa, Phys. Rep. {\bf 407}, 205 (2005).
\vfill\eject
\end{thebibliography}
\end{document}